\documentclass[10pt,%
aps,
pra,
superscriptaddress,
amsmath,amssymb,
twocolumn,
]{revtex4-2}
\usepackage{nameref}
\usepackage[colorlinks,
 linkcolor = black,
 urlcolor = black,
 anchorcolor = black,
 filecolor = black,
 ]{hyperref}
\usepackage{amsmath}
\usepackage{titlesec}
\usepackage{graphicx}% Include figure files
\usepackage{dcolumn}% Align table columns on decimal point
\usepackage{bm}% bold math
\usepackage{rotating}
\usepackage{mathrsfs}
\usepackage[dvipsnames]{xcolor}

\begin{document}
\titlespacing{\section}
 {0pt} %left
 {10pt} % before
 {0pt} % after
\titleformat*{\section}{\Large\bfseries\raggedright}
\titlespacing{\subsection}
 {0pt} % left
 {15pt} % before
 {5pt} % after
\titleformat{\subsection}[block] 
 {\bfseries}
 {}
 {0pt}
 {}

\title{PAEMS: Precise and Adaptive Error Model for Superconducting Quantum Processors}% Force line breaks with \\

\author{Songhuan He}
\affiliation{School of Electronic Science and Engineering, University of Electronic Science and Technology of China, Chengdu, Sichuan, China}
\author{Yifei Cui}
\affiliation{School of Electronic Science and Engineering, University of Electronic Science and Technology of China, Chengdu, Sichuan, China}
\affiliation{University of California, Los Angeles, USA}
\author{Bo Liu}
\thanks{Email: liubo08@nudt.edu.cn}
\affiliation{College of Advanced Interdisciplinary Studies, National University of Defense Technology, Changsha, Hunan, China}
\author{Kai Guo}
\thanks{Email: guokai07203@hotmail.com}
\affiliation{Institute of Systems Engineering, Academy of Military Science, Beijing, China}
\author{Cheng Wang}
\thanks{Correspondence to: wangch87@uestc.edu.cn}
\affiliation{School of Electronic Science and Engineering, University of Electronic Science and Technology of China, Chengdu, Sichuan, China}

\date{\today}

\begin{abstract}
Superconducting quantum processor units (QPUs) are incapable of producing massive datasets for quantum error correction (QEC) because of hardware limitations. Thus, QEC decoders heavily depend on synthetic data from qubit error models. Classic depolarizing error models with polynomial complexity present limited accuracy. Coherent density matrix methods suffer from exponential complexity $\propto O(4^n)$ where $n$ represents the number of qubits. This paper introduces PAEMS: a precise and adaptive qubit error model. Its qubit-wise separation framework, incorporating leakage propagation, captures error evolvements crossing spatial and temporal domains. Utilizing repetition-code experiment datasets, PAEMS effectively identifies the intrinsic qubit errors through an end-to-end optimization pipeline. Experiments on IBM's QPUs have demonstrated a 19.5$\times$, 9.3$\times$, and 5.2$\times$ reduction in timelike, spacelike, and spacetime error correlation, respectively, surpassing all of the previous works. It also outperforms the accuracy of Google’s SI1000 error model by 58$\sim$73\% on multiple quantum platforms, including IBM’s Brisbane, Sherbrooke, and Torino, as well as China Mobile’s Wuyue and China Telecom’s Tianyan.
\end{abstract}
\maketitle

\section*{Introduction}

\begin{figure*}
\includegraphics[width=1\textwidth]{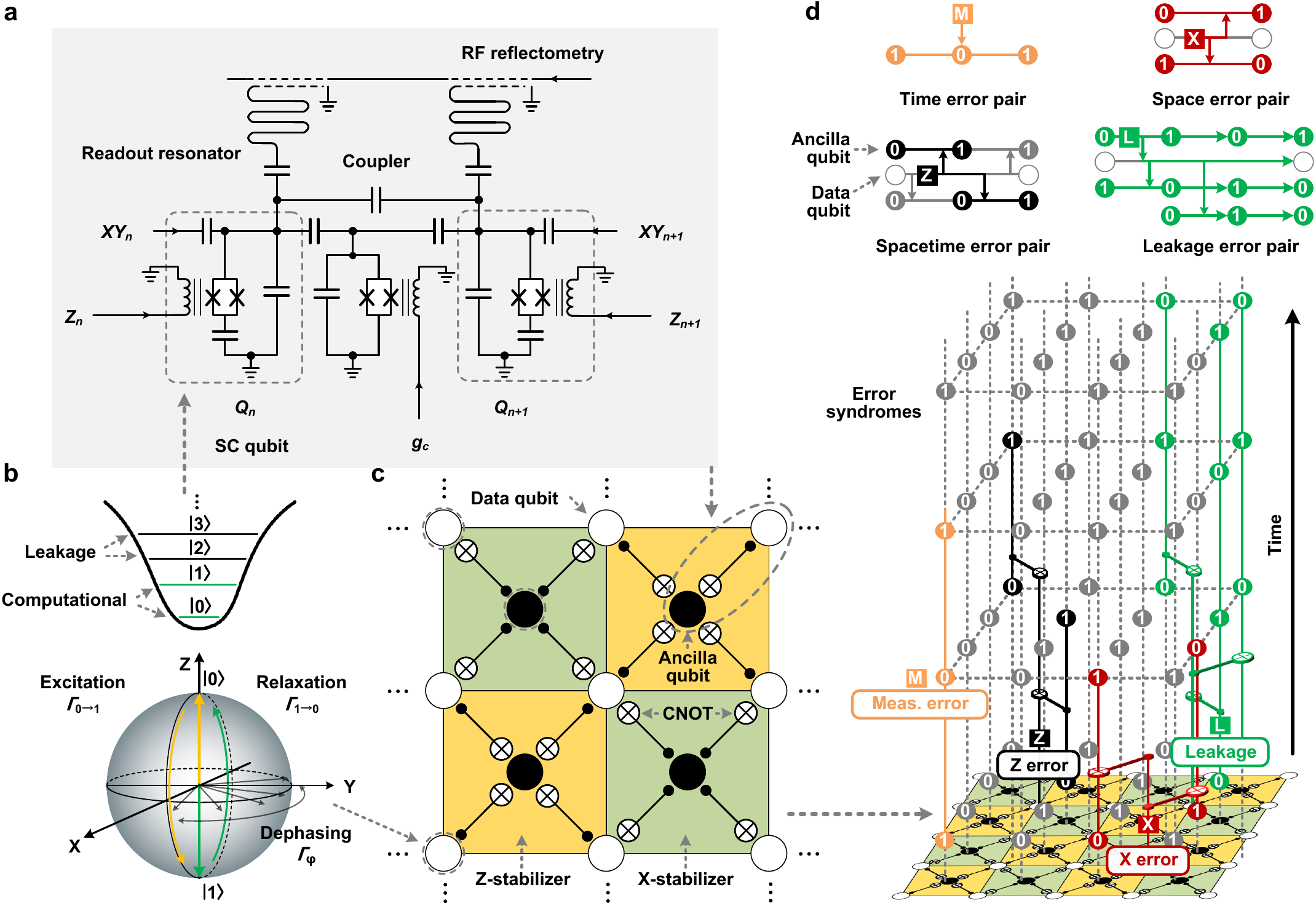}
\caption{\label{fig1}Errors of a superconducting qubit array. 
\textbf{a} Two coupled transmon qubits. Each qubit employs XY RF pulses and Z flux pulses for the single-qubit gates ($e.g.$ Hadamard gate). A tunable coupler enables the two-qubit gates ($e.g.$ CNOT, CZ). The RF dispersive reflectometer measures the qubit state. 
\textbf{b} Error mechanisms of a single qubit: besides the excitation ($\Gamma_{0\rightarrow1}$), relaxation ($\Gamma_{1\rightarrow0}$), and dephasing ($\Gamma_{\phi}$), leakage beyond the computational subspace is critical. These errors come from unwanted transition and phase shift in a multi-level system. 
\textbf{c} A surface code topology with data and ancilla qubits. Z-stabilizers (yellow) and X-stabilizers (green) are implemented through coupled qubit pairs of \textbf{a}. 
\textbf{d} Error syndromes reflect the origin of qubit errors: Z-stabilizers detect Pauli-X errors, while X-stabilizers detect Pauli-Z errors. Different errors and locations lead to varying time, space, and spacetime error syndromes and the corresponding error pairs. In particular, a leakage event can persist across multiple rounds and propagate through entangling gates, generating multiple correlated error pairs spanning over space and time. 
}
\end{figure*}

Superconducting (SC) quantum processors have been one of the leading platforms of quantum error correction (QEC) \cite{calderbank1996good, fowler2012surface} and quantum error mitigation \cite{king2022coherent, giurgica2020digital}, due to the adequate number of qubits ($10^2\sim10^3$), mutually-coupled 2-D planar configuration, and mature quantum-classic interface. Recently, successive breakthroughs have been observed, such as logical qubit overcoming additional errors from up-scaling \cite{google2023suppressing}, below-threshold surface code memories on Google's SC quantum processor "Willow" \cite{google2025quantum}, color code implementations with logical error suppression scaling \cite{lacroix2025scaling}, and SC quantum processor demonstrating random sampling \cite{PhysRevLett.134.090601}. Quantum processors with up to 1121 SC qubits have been announced \cite{abughanem2024ibm, cas_2024_tianyan504}. Consequently, ref. \cite{Mohseni_2025} envisions a road map towards millions of SC qubits for a practical quantum computer. However, bottle-necked by the non-ideality (e.g. qubit variation, slow calibration, and control $\&$ readout infidelity), the QEC has been validated only on small scale (5$\sim$105 qubits) so far with insufficient statistics. In contrast, for topological-code based QEC \cite{dennis2002topological} (e.g. surface code, color code), a fault-tolerant quantum computer demands $\sim 10^3$ physical qubits for 1 logical qubit \cite{fowler2012surface}. Accordingly, an accurate, low latency QEC decoder is indispensable to identify the errors from the sophisticated syndromes, before applying the feedback correction. 
\par
To this end, minimum weight perfect matching (MWPM) has been adopted as the baseline of QEC decoders \cite{fowler2012surface}. However, its accuracy is hindered by the non-uniformity, noise correlation, and leakage of qubits. A computational complexity of $\propto O(n^3)$ also limits the scalability of MWPM decoders. To deal with these issues, MWPM-BP \cite{caune2023belief} and MWPM-Corr \cite{yuan2022modified} are proposed. Alternatively, union-find decoders \cite{delfosse2022toward} enjoy a reduced complexity of $\propto O(n)$, but suffer from a precision penalty. Subsequently, riverlane-CC \cite{barber2025real} and union-find BP \cite{delfosse2022toward} achieve a better balance between accuracy and efficiency. Furthermore, neural network (NN) QEC decoders are promising. Google's transformer-based AlphaQubit \cite{bausch2024learning} achieves a logical error rate (LER) of 2.748±0.015\% in code distance of $d$=5, compared to 2.915±0.016\% for tensor network decoders \cite{google2023suppressing, bravyi2014efficient}. USTC's GraphQEC based on graph neural network (GNN) achieves an LER of $9.55 \times 10^{-5}$ for a QLDPC code with $d$=12, which is 18$\times$ better than the state-of-the-arts \cite{hu2025efficient}. Other NN decoders also reported promising results \cite{varbanov2025neural, overwater2022neural, lange2025data}. Their real-time hardware implementations are summarized in \cite{battistel2023real}.
\par
However, training of these QEC decoders demands massive error datasets, which cannot be directly obtained from quantum processors. This is due to the limited sizes of qubits, non-idealities, and high cost. Therefore, a precise qubit error model is indispensable to capture the realistic error correlations and spatiotemporal dynamics of large-scale qubit array. For instance, AlphaQubit adopts synthetic data up to 2$\times 10^9$ samples for pretraining, generated by the detector error model (DEM) and the SI1000 depolarizing noise model \cite{bausch2024learning}. AlphaQubit also requires $10^8$ additional samples for fine-tuning, coming from the Sycamore experimental dataset and the Pauli+ noise model. USTC's GraphQEC \cite{hu2025efficient} also adopts a depolarization noise model \cite{gidney2021stim} for pretraining, and Google's Sycamore dataset for fine-tuning. It should be noted that Google’s Sycamore dataset only contains 6.5$\times 10^6$ surface code shots  \cite{google_quantum_ai_team_2022_6804040}. However, firstly, SC qubits, as shown in (Fig. \ref{fig1}\textbf{a}), are subjected to energy relaxation, dephasing, and leakage to non-computational states (Fig. \ref{fig1}\textbf{b})  \cite{wallman2016noise,salari2024microwave, resch2021benchmarking}, while exhibiting device inhomogeneity and time-varying event rate. Quantum operations, such as initialization, single-qubit gates, two-qubit gates, and readout, also exhibit infidelity with spatiotemporal variation (Fig. \ref{fig1}\textbf{c}). Those error mechanisms create sophisticated time, space, spacetime and leakage error syndromes, as shown in Fig. \ref{fig1}\textbf{d}. It poses significant challenges on modeling accuracy \cite{google2023suppressing, google2025quantum}. Secondly, a comprehensive qubit error model shall also be adaptive over different quantum processors. In addition, there is a fundamental trade-off between accuracy and complexity. Through propagating $2^n\times2^n$ density matrices for each gate, coherent density matrix methods \cite{greenbaum2017modeling, feng2016estimating, barnes2017quantum, bravyi2018correcting, o2017density} enjoy high accuracy, but burden a computational complexity of $\propto O(4^n)$. Thus, it typically works for small-scale QPUs with $<$ 20 qubits \cite{bravyi2018correcting}. Alternatively,  based on the Gottesman-Knill theorem \cite{aaronson2004improved, gottesman1998heisenberg}, by inserting depolarizing channels at each step, the stochastic Pauli models \cite{katabarwa2015logical, tomita2014low, geller2013efficient} present a computational complexity of $\propto O(n^2)$. However, the assumption of uniform qubits crossing the chip is problematic. Even Google's SD6 and SI1000\cite{gidney2021fault} only differentiate errors of individual quantum operations ($e.g.$ gate, measurement, reset), ignoring the spatiotemporal variations \cite{google2023suppressing}. Consequently, trained on the synthetic data, NN decoders present a 15$\sim$25\% accuracy degradation \cite{varsamopoulos2019comparing, bausch2024learning}.
\par
This paper introduces PAEMS: a depolarizing error model with high precision and cross-platform adaptivity. To deal with the device non-uniformity, it adopts qubit-specific parameters to model the decoherence, gate infidelity, and state preparation and measurement (SPAM) errors. It also incorporates the leakage and seepage rate for error propagation in the spatiotemporal domains. Additionally, to capture the platform dependent error behaviors, experiment datasets are directly sampled on the target QPUs using a repetition-code topology. Then, initialized by the qubit calibration, PAEMS is trained on the experiment datasets through the covariance matrix adaptation evolution strategy (CMA-ES), which avoids double-counting of overlapping error sources and achieves cross-platform adaptivity. In the experiments on IBM's Brisbane, Sherbrooke, and Torino QPUs \cite{ibm_quantum}, accurate modeling of a 30-round repetition-code with 21-qubits achieves a $19.5\times$, $9.3\times$, and $5.2\times$ reduction in timelike, spacelike, and spacetime correlation differences, respectively, compared with the state-of-the-arts. Through IBM’s Brisbane, Sherbrooke, and Torino QPUs, as well as China Mobile’s Wuyue \cite{China_Mobile_Wuyue} and China Telecom’s Tianyan QPUs \cite{China Telecom_Tianyan}, the cross-platform adaptivity of PAEMS is validated using single-round repetition-code experiments with 5$\sim$21 qubits. In all cases, PAEMS achieves a 58$\sim$73\% reduction in total variation distance (TVD) compared with Google’s SI1000 error model. Lastly, PAEMS scales friendly with the expansion of qubit array, achieving a computational complexity of $\propto O(n^2)$.
\par

\section*{Results}
\subsection*{Spatiotemporal Heterogeneity within Multi-Round Repetition Code}

\begin{figure*}
\includegraphics[width=1\textwidth]{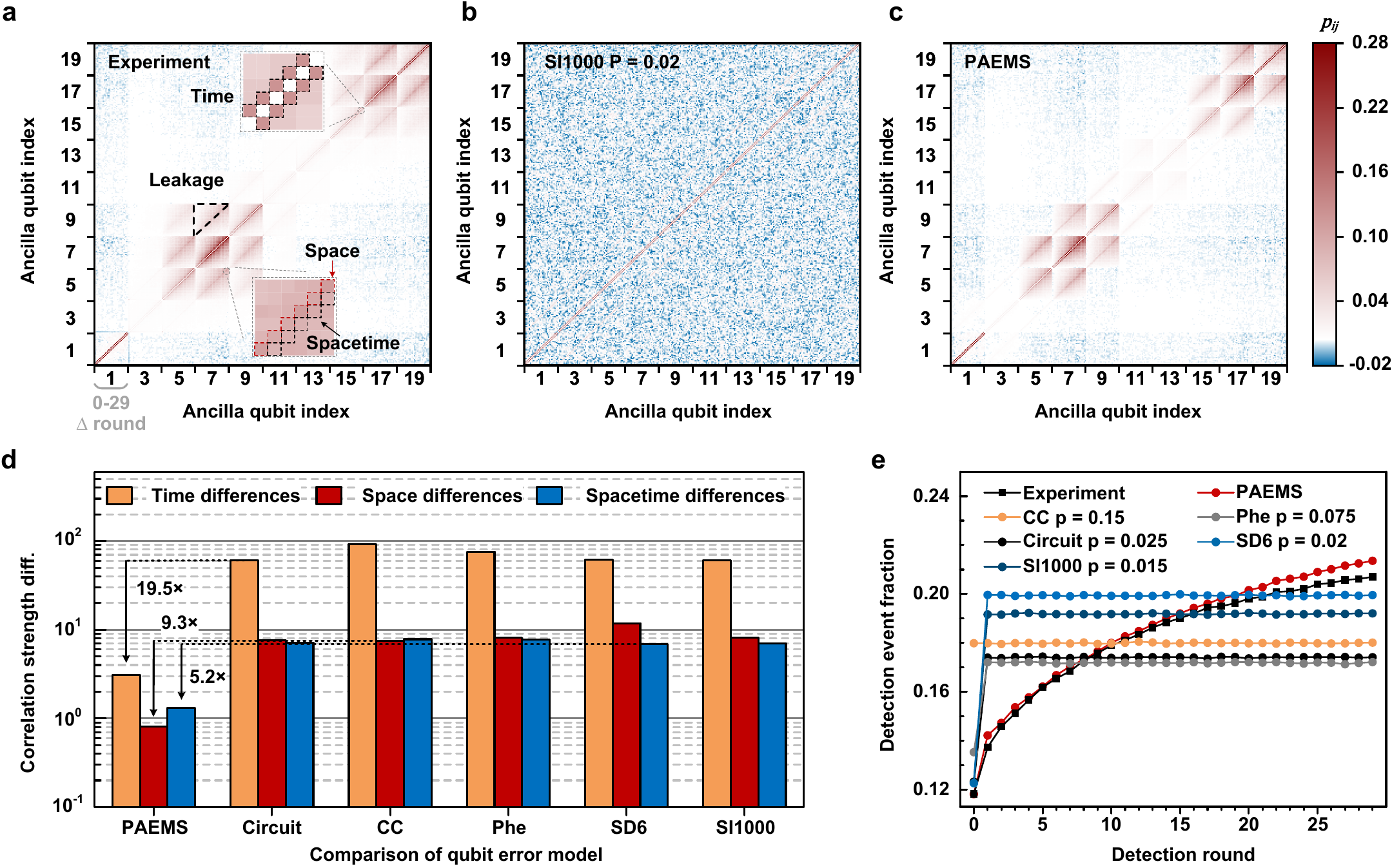}
\caption{\label{fig2}Repetition code experiment with 21 qubits and 30 detection rounds.
\textbf{a$\sim$c} Correlation matrices of experiment, SI1000 model with $p$=0.02 and PAEMS model. Different element sets of the symmetric correlation matrix correspond to different errors, where $p_{ij}$ denotes the two-point correlations between detection events. 
\textbf{d} Correlation strength differences between the experiment and the qubit error models. PAEMS, Circuit, Code-capacity (CC), Phenomenological (Phe), SD6, and SI1000 models are simulated in time, space, and spacetime differences. 
\textbf{e} Detection event fraction indicating the event intensity over 30 successive rounds of repetition code. The experiment results, PAEMS, Circuit ($p$=0.025), CC ($p$=0.15), Phe ($p$=0.075), SD6 ($p$=0.02), and SI1000 ($p$=0.015) models are plotted.}
\end{figure*}

As illustrated in $Methods$, repetition code preserves the local entanglement and dominant error-propagation of surface code, while remaining experimental feasibility on mainstream quantum platforms. Thus, it is employed for the model evaluation, as a reduced yet structurally relevant proxy for surface codes. A 30-round repetition code is adopted here to probe the spatiotemporal error correlations, built on IBM's QPUs with 21 qubits. The measured correlation matrix of IBM's Torino QPU is shown in Fig. \ref{fig2}\textbf{a}. Here, $p_{ij}$ quantifies the correlation between two detection events, spanning across the ancilla qubits and the repetition rounds \cite{google2021exponential}. It can be observed that: First, the measured detection-event correlations are not uniform. For timelike correlations, ancilla qubit 1 exhibits a relatively high correlation of $0.335 \pm 0.012$, whereas ancilla qubit 3 only shows $0.015 \pm 0.002$, differing by a factor of $22.3$. This reveals substantial difference in measurement-error probabilities of qubits. Second, leakage errors lead to distinct correlation sectors that fall into 3 categories: timelike, spacelike, and spacetime correlations. The measured correlation of ancilla qubit 7 increases monotonically from $0.096$ for the $1^{st}$ round to $0.465$ for the $29^{th}$ round, indicating an accumulation of leakage-induced errors. Ancilla qubit 17 also exhibits similar leakage-related accumulation. In the worst case, the disparity among correlations of the same category can reach up to $10^{2}$, which hinders the QEC decoders with uniform qubit assumption.
\par
Next, the 30-round repetition code is simulated by Google's depolarizing error model SI1000 \cite{gidney2021fault} with error rate of $p$=0.02, as shown in Fig. \ref{fig2}\textbf{b}. Since SI1000 assumes identical qubits, a high degree of uniformity is observed for all of the timelike ($0.143 \pm 0.013$), spacelike ($0.043 \pm 0.011$), and spacetime ($0.011 \pm 0.010$) correlations. It also produces no entries of two-point correlation $p_{ij}$ corresponding to leakage correlation sectors. In addition, many of the extracted $p_{ij}$ are negative, which reflects the statistical independence imposed by the Markovian assumptions. Accordingly, the correlated error mechanisms beyond single-component processes are largely absent, causing pairwise correlation estimates to vanish or become negative. In contrast, platform-specific qubit parameters are extracted first for the initialization of PAEMS model (see $Methods$). The same 30-round repetition-code experiment is then simulated by PAEMS in Fig. \ref{fig2}\textbf{c}, which reproduces the correlation distribution closer to the experiment. Specifically, the timelike correlation of ancilla qubit 1 is $0.337 \pm 0.008$, while that of ancilla qubit 3 is $0.015 \pm 0.001$. In addition, the timelike correlation of ancilla qubit 7 increases from $0.127$ of the $1^{st}$ round to $0.476$ of the $29^{th}$ round, capturing the error-accumulation observed experimentally. The simulation also reproduces the distinct leakage-specific correlation sector of the experiment. Furthermore, the discrepancy between the model and experiment can be quantified by the correlation strength difference. Then, PAEMS and the other state-of-the-art depolarizing qubit error models (Circuits \cite{fowler2012surface}, CC \cite{dennis2002topological}, Phe \cite{wang2009threshold}, SD6 and SI1000 \cite{gidney2021fault}, evaluated at their optimal physical error rate) are compared in Fig. \ref{fig2}\textbf{d}. It can be observed that PAEMS achieves an overall accuracy improvement of $19.5\times$, $9.3\times$, and $5.2\times$ in timelike, spacelike, and spacetime correlation differences, respectively. Lastly, in Fig. \ref{fig2}\textbf{e}, the detection-event fraction across 30 detection rounds predicted by PAEMS follows the experiment closely, with a root-mean-square (RMS) error of $3.4 \times 10^{-3}$ only. In contrast, due to Markovian-based structure, simulated detection-event fraction of the other models quickly saturates after the $1^{st}$ round, failing to reproduce the experiments. More data can be found in the Supplementary.

\subsection*{Cross-Platform Adaptivity with Single-Round Repetition Code}

\begin{figure*}
\centering
\includegraphics[width=1\linewidth]{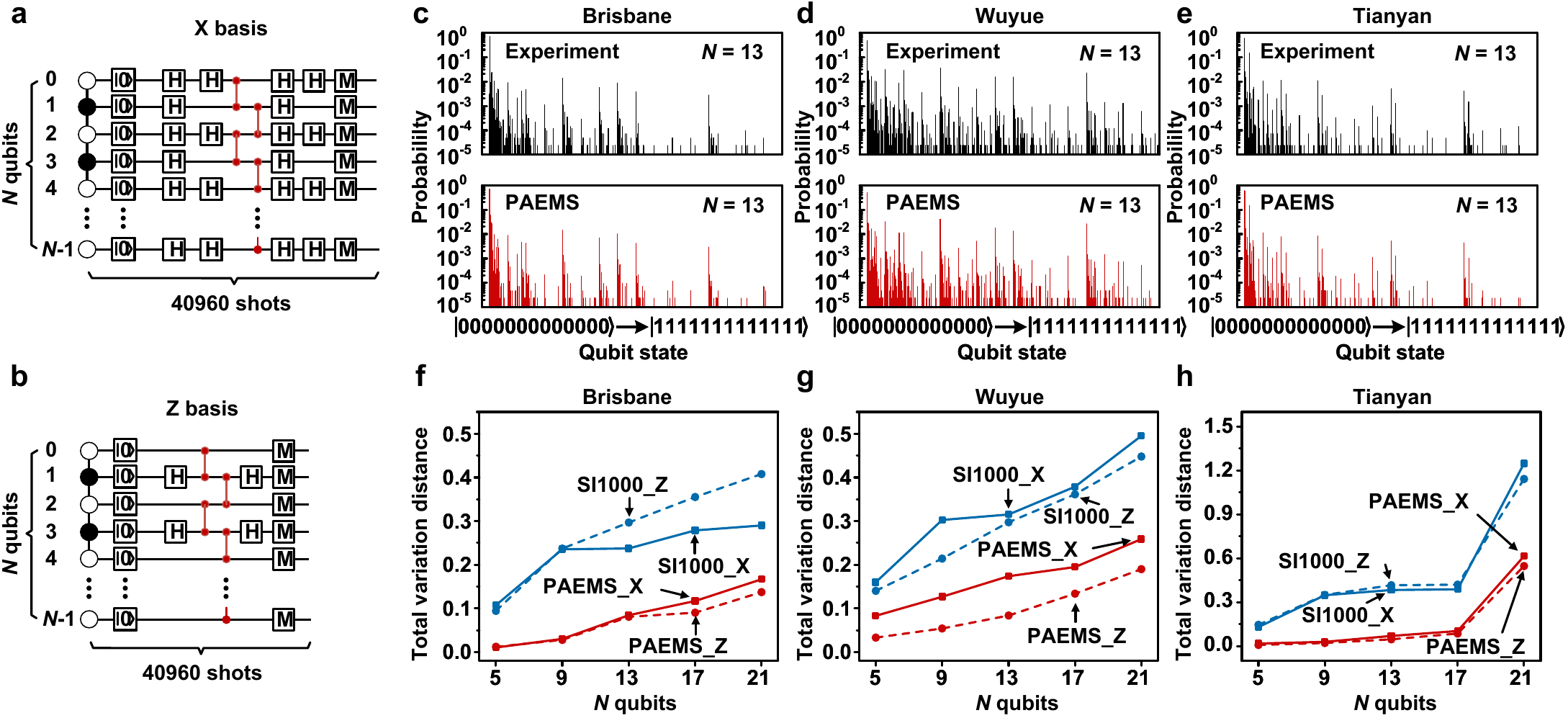}
\caption{\label{fig3}Cross-platform adaptivity with single-round repetition code. 
\textbf{a} X basis and \textbf{b} Z basis repetition-codes with $N$ = 5, 9, 13, 17, 21 qubits. 40960 shots are measured for each case. 
\textbf{c$\sim$e} Measured probability distributions of X basis repetition code with $N=13$ on Brisbane, Wuyue, and Tianyan, respectively, compared with PAEMS model.
\textbf{f$\sim$h} Total variation distance (TVD) between experiment and model prediction. PAEMS and SI1000 models are compared in both X and Z basis. SI1000 model is simulated with the optimal $p$.
}
\end{figure*}

To ensure consistency across heterogeneous quantum platforms, quantum error correction experiments are conducted across multiple platforms QPUs. Owing to hardware limitations, single-round repetition codes are adopted as the comparison baseline. As shown in Fig. \ref{fig3}, the experimental results are obtained simultaneously from IBM's Brisbane, China Mobile’s Wuyue, and China Telecom‘s Tianyan QPUs. Two different single-round repetition codes, including the X basis (Fig. \ref{fig3}\textbf{a}) and the Z basis (Fig. \ref{fig3}\textbf{b}), are employed with a qubit size of $N$=5, 9, 13, 17, and 21. Quantum state discrimination then produces qubit readouts over $2^{N}$ computational states and the output-state probability distribution. 40960 shots for each circuit have been measured for a meaningful statistics. In the experiment, the qubits are initialized into all-zero states. Thus, if it is error free, the outcome shall also be all-zero. Accordingly, after the single-round repetition code, a higher probability of all-zero state implies a better fidelity.
\par
For the X basis repetition code with $N$=5, the measured all-zero state probabilities are $0.922$ for Brisbane, $0.769$ for Wuyue, and $0.924$ for Tianyan, respectively. However, as shown in Fig. \ref{fig3}\textbf{c}$\sim$\textbf{e}, the measured all-zero state probability decreases to $0.709$ for Brisbane, $0.477$ for Wuyue and $0.602$ for Tianyan under $N$=13. The decreasing is more significant as $N$ increases. Then, the all-zero state probabilities simulated by PAEMS under $N$=13 are $0.696$ for Brisbane, $0.512$ for Wuyue, and $0.604$ for Tianyan, respectively, which are close to the experiment. To quantify the discrepancy between experiment and simulation, the total variation distance (TVD) is used, which is evaluated over the full $2^{N}$-dimensional state space of the repetition code. As shown in Fig. \ref{fig3}\textbf{f}$\sim$\textbf{h}, on all of the Brisbane, Wuyue and Tianyan QPUs, TVDs of both of the $X$ and $Z$ basis repetition code simulated by PAEMS are lower than that of SI1000 from $N=5\sim21$. Specifically, all TVD (aggregated over $N=5\sim21$ and both measurement bases) of PAEMS model is $69.4\%$ lower on Brisbane, $58.1\%$ lower on Wuyue, and $68.9\%$ lower on Tianyan, compared with SI1000. Additional data and the definition of TVD can be found in the Supplementary.
\par
\section*{Methods}
\subsection*{Circuit-level Error Modeling Framework of PAEMS}

\begin{figure*}
\centering
\includegraphics[width=1\linewidth]{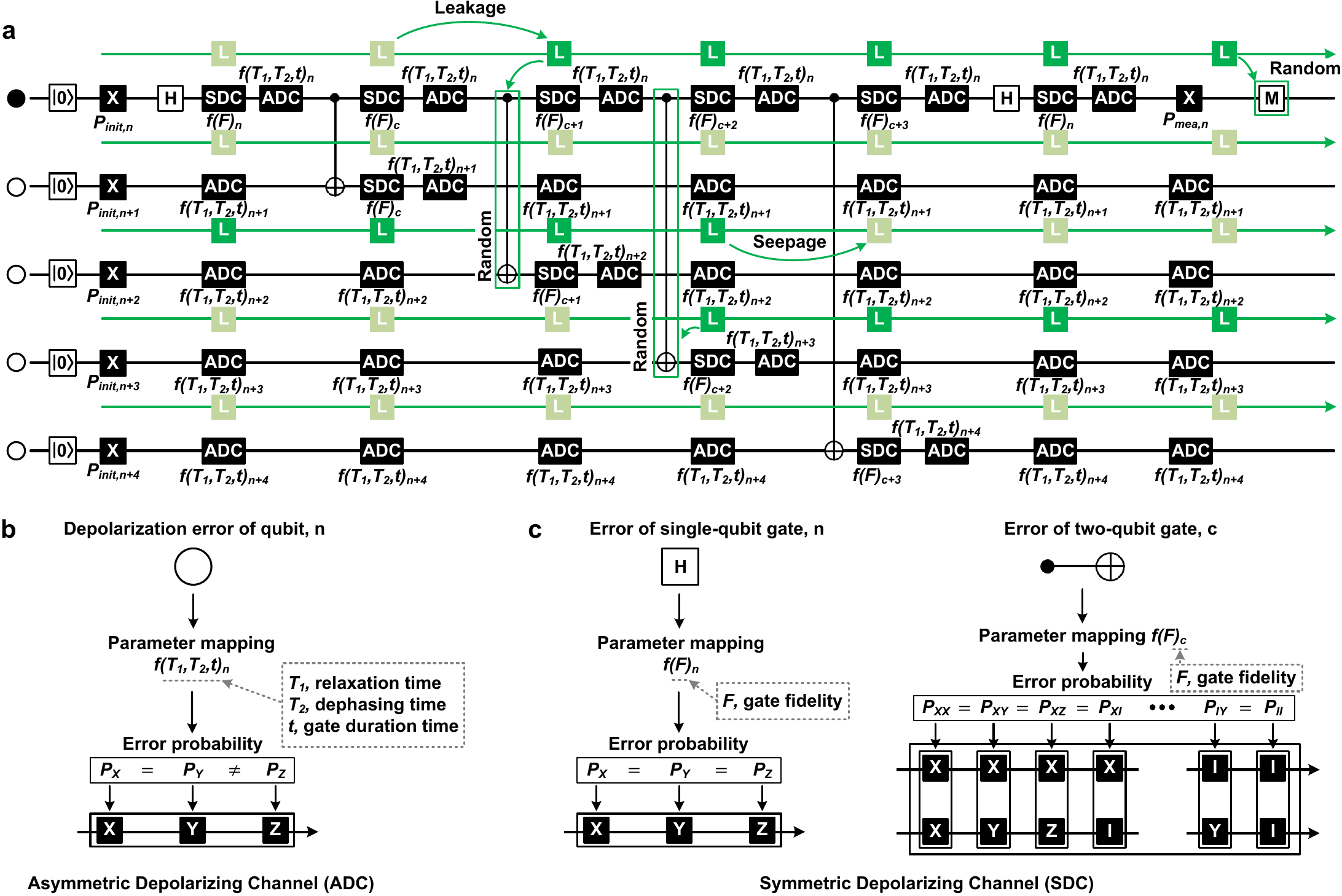}
\caption{\label{fig4}Framework of PAEMS qubit error model. 
\textbf{a} Circuit-level representation of error channels, which is parameterized per qubit and explicitly resolved in the time domain. Decoherence errors are modeled by asymmetric depolarizing channels (ADC). Gate errors are modeled by symmetric depolarizing channels (SDC). State preparation and measurement (SPAM) errors are modeled by stochastic Pauli-$X$ channels.
Leakage and seepage processes evolve in parallel with Pauli error channels, and their error propagation is simulated by randomizing the outcomes of two-qubit gate operations and measurements to $\lvert 0\rangle$ or $\lvert 1\rangle$ with equal probability.
\textbf{b} Parameter mapping of ADC channels, where decoherence errors of qubit $n$ are initialized by the relaxation time $T_1$, dephasing time $T_2$, and gate or idle duration time $t$ from qubit calibration. It leads to stochastic Pauli error channels with realistic non-uniformity.
\textbf{c} Parameter mapping of SDC channels for single-qubit and two-qubit gates, parameterized by gate fidelities $f(F)_n$ and $f(F)_c$, respectively. It results in stochastic Pauli error channels with equal probabilities. $n$ is the qubit index, while $c$ labels the coupler index.}
\end{figure*}

PAEMS is formulated as a circuit-level stochastic error model of superconducting qubits calibrated through a data-driven optimization pipeline. It reflects a deliberate trade-off between physical realizability and computational scalability. In contrast, classic density-matrix methods capture the coherent dynamics with high fidelity, but their application is hindered by the exponential complexity and the demands of highly accurate noise parameters. The Pauli+ model \cite{google2023suppressing} alleviates the computational efficiency issue by mapping density-matrix noise channels to generalized Pauli channels. However, it does not necessarily yield error statistics that match experimental observations. DEM \cite{gidney2021stim} operate purely on the detection-event statistics, which do not explicitly encode how underlying physical error processes propagate through quantum circuits to generate those events. In comparison, a balance of model accuracy and physical interpretability is achieved by PAEMS. It takes a circuit-level description of underlying noise processes, enabling high-fidelity reproduction of detection-event statistics, while preserving a transparent connection to hardware-level error mechanisms. Fig. \ref{fig4}\textbf{a} illustrates the PAEMS framework. Each operation is modeled as an ideal quantum gate, followed by parameterized stochastic error channels. This allows natural error propagation across both of the space and time domain, also maintaining high computational efficiency. More importantly, while the existing circuit-level error models relying on globally averaged and phenomenological error rates, PAEMS resolves errors at the granularity of individual qubits, couplers, and circuit layers. This enables effective modeling of the spatial heterogeneity and the accumulation of temporal error. To this end, PAEMS adopts a physics-informed parameterization based on platform-calibrated quantities for each qubit.
\par
Next, the gate fidelity reflects errors from the imperfection of gate control and the decoherence. Therefore, an aggregated gate fidelity parameter often leads to double-counting and obscuring the physical origin of errors. To address this issue, an asymmetric depolarizing channel (ADC) of PAEMS separates decoherence-induced errors from gate infidelity using the calibrated relaxation time $T_1$, dephasing time $T_2$, and operation duration $t$ \cite{tomita2014low}, as illustrated in Fig. \ref{fig4}\textbf{b}. In addition, after separating the decoherence-errors, symmetric depolarizing channels (SDC), parameterized by the gate fidelity $F$, are adopted to model the  errors of single and double qubit gates (Fig. \ref{fig4}\textbf{c}). Furthermore, for the state preparation and measurement (SPAM) errors, the state preparation and reset, following the intermediate measurements, general present distinct infidelity for a given physical qubit due to the platform-specific implementations. To account for this asymmetry, PAEMS assigns two distinct state-preparation error probabilities ($P_{init,i}$ and $P_{mea,i}$ in Fig. \ref{fig4}\textbf{a}) to each qubit, corresponding to initialization at the beginning of the circuit and reset after measurement during repeated QEC cycles. This treatment preserves a clear correspondence between model parameters and underlying physical processes while maintaining circuit-level efficiency.
\par
Finally, in a quantum circuit, a qubit can stochastically transition from the computational subspace to a leakage state. Once leaked, the qubit remains outside of the computational subspace, and subsequently affects the circuit operations according to predefined leakage rules(randomizing two-qubit gate and measurement operations). Eventually, the leaked qubit will return to the computational subspace through a seepage process. PAEMS models the leakage by introducing a leakage and a seepage probability for each qubit. These leakage and seepage processes are applied at each relevant circuit location as shown in Fig. \ref{fig4}, enabling propagation across multiple rounds of quantum circuits.

\subsection*{Experiment Dataset and Training of PAEMS}

\begin{figure*}
\centering
\includegraphics[width=1\linewidth]{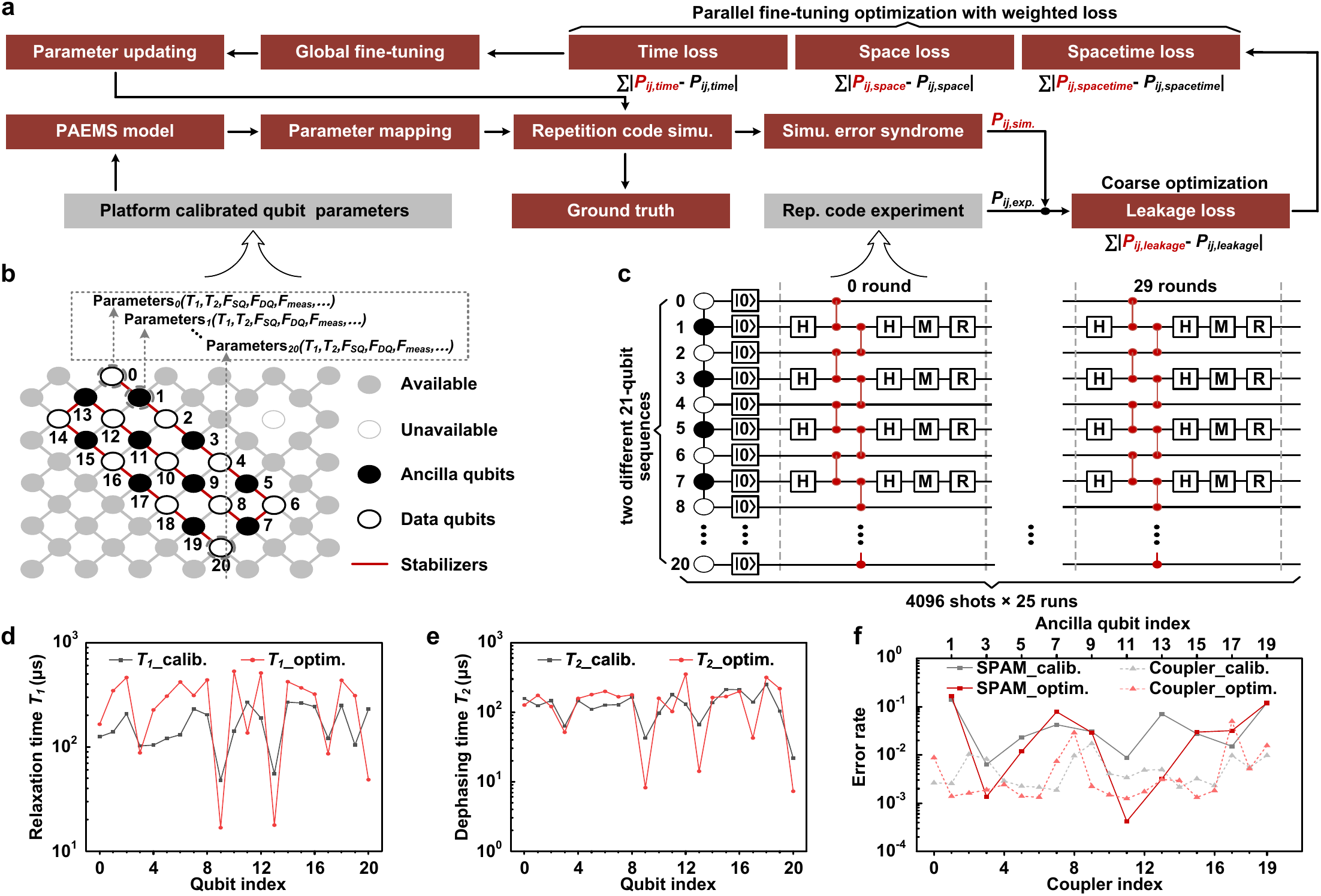}
\caption{\label{fig5}Training of PAEMS model. 
\textbf{a} End-to-end workflow of model training. The experiment inputs (gray boxes) include the qubit calibration and the sampled repetition-code datasets. The training pipeline (red boxes) starts with the parameter mapping, repetition-code simulation, and error-syndrome generation. Model optimization is performed in three stages: coarse optimization of leakage parameters, parallel fine-grained optimization using time, space, and spacetime loss terms, and final global fine-tuning. All stages employ the CMA-ES algorithm.
\textbf{b} The 21-qubit repetition code topology for dataset generation, with platform-calibrated parameter collection performed on each physical qubit and coupler.
\textbf{c} Quantum circuit of multi-round 21-qubit repetition code, including 30 detection rounds with 4096 shots per run, repeated 25 times. Two different qubit sequences are tested on each quantum chip.
\textbf{d,e} Comparison between platform-calibrated and optimized decoherence times for the Torino processor, including the relaxation time $T_1$ and dephasing time $T_2$ of the 21 tested qubits.
\textbf{f} Comparison between platform-calibrated and optimized SPAM error rates of the ancilla qubits, and the error rate of couplers, from the Torino processor.}
\end{figure*}

An end-to-end optimization workflow is adopted by PAEMS, as illustrated in Fig. \ref{fig5}\textbf{a}. Platform calibrated qubit parameters are first used to initialize the PAEMS model. A 21 qubit repetition code scheme with two distinct qubit layouts on the same QPU is then employed, as shown in Fig. \ref{fig5}\textbf{b} and Supplementary. The 30 round circuit in Fig. \ref{fig5}\textbf{c} is executed for 25 independent runs with 4096 shots per run. The resulting measurement data are processed to extract detection events and correlation statistics. These statistics are used to optimize the PAEMS parameters. Because a leakage event can impact multiple detection event correlations $p_{ij}$, and bias the inference of other parameters, leakage related parameters are treated separately and optimized first. Since the leakage parameters are not included in the calibration, they are initialized to zero, and inferred through a dedicated coarse optimization stage subsequently. Next, the other error parameters, including the relaxation time $T_1$, dephasing times $T_2$, fidelity of quantum gates, and readout fidelity of ancilla qubits, are optimized. They contribute to temporal, spatial, and spatiotemporal correlations with different relative weights. A parallel optimization scheme is then adopted, where different correlation deviations are incorporated with distinct loss weights. In addition, the detection event fraction is used to extract state preparation errors. Finally, a global refinement of all parameters is performed. For numerical stability and computational efficiency, optimization is carried out on the correlation data averaged over 25 runs first. It is then followed by detailed optimization for individual run to account for the run-to-run variations.
\par
Fig. \ref{fig5}\textbf{d}$\sim$\textbf{f} compare the qubit calibration of Torino QPU with the optimized parameters by PAEMS. A divergence of qubit parameters with a factor of 3 to 5 is observed, which is notable in the relaxation and dephasing times $T_1$ and $T_2$ of qubit 9, 10, 12, and 13. For example, qubit 9 shows a parameter variation from $T_1 = 47.9~\mu$s and $T_2 = 42.3~\mu$s to $T_1 = 16.7~\mu$s and $T_2 = 8.1~\mu$s after the optimization. In addition, Fig. \ref{fig5}\textbf{f} shows that the state preparation error probability of qubit 11 substantially reduced from $8.79\times10^{-3}$ to $4.21\times10^{-4}$ after the optimization, while the SPAM and coupler related parameters of other qubits also exhibit varying degrees of adjustment (More data can be found in the Data availability). Such differences are mainly due to: (1) the calibration protocols and instrumentations differ by platforms \cite{gambetta2017building, mckay2017efficient}. The calibration are also not performed in real time. (2) the calibration does not take the following facts into consideration: coupling of qubits, repeated measurement, and reset operations, which are intrinsic to surface code and repetition code. Therefore, it leads to systematic deviations \cite{gambetta2017building, mckay2017efficient, google2023suppressing, google2021exponential}. (3) the qubit parameters drift over time under low frequency noise and parasitic two level systems (TLS) \cite{muller2019towards, klimov2018fluctuations}. To this end, qubit parameters within a single QPU differing by $10^1$ to $10^2$. Individual qubit parameters may drift by a factor of $5\times$.
\par
As a special note, for single round repetition code experiments, correlation based metrics are not available due to the absence of temporal and spatial error accumulation. In this case, core qubit parameters are normalized and jointly optimized within a unified optimization procedure. Leakage processes are not considered in the single round setting, and the leakage and seepage probabilities are therefore fixed to zero. The optimization is performed using the TVD between experimentally measured and simulated output state probability distributions as the objective metric.

\section*{Discussion}
Training of advanced QEC decoders, such as MWPM-BP and neural-network-based approaches, demands massive and accurate synthetic error datasets from qubit error models. However, previous models heavily rely on uniform, Markovian, or phenomenological assumptions, which fail to reproduce the correlated detection-event statistics in real quantum circuits. In the meanwhile, as the noise models suppress correlations or average over qubit-specific behavior, they systematically misrepresent the effective error landscape encountered during circuit execution. It leads to degraded decoder performance on hardware. As shown in the multi-round repetition code experiments of Fig.\ref{fig2}, these models underestimate spatial heterogeneity, temporal error accumulation, and leakage-induced correlations. Therefore, mismatch of simulated and measured qubit error syndromes can be observed. 
\par
In PAEMS, by parameterizing individual qubit and coupler, modeling error propagation and leakage, and training with the qubit calibrations and experiment datasets, the resulting noise statistics faithfully reproduce the measurement. In Fig. \ref{fig2}, PAEMS effectively lowers the discrepancies of correlation strength by factors of $19.5\times$, $9.3\times$, and $5.2\times$ in the timelike, spacelike, and spacetime sectors, respectively. Furthermore, in the single-round repetition-code of Fig. \ref{fig3}, PAEMS also exhibits a $58\text{--}73\%$ reduction in TVD compared to SI1000, across multiple quantum platforms. With a computational complexity of $O(n^2)$, PAEMS enables scalable and reliable training of QEC decoder. It also opens the door of physically interpretable qubit error models, which can be iteratively refined alongside QEC decoders. 
\par
At present, PAEMS does not explicitly model additional noise mechanisms such as coherent errors and qubit crosstalk induced by simultaneous gate operations. Future work will extend PAEMS to incorporate effective circuit-level models of crosstalk and coherent noise. The framework will be validated using experiments with quantum error correction codes beyond repetition codes, such as surface codes and color codes. In addition, PAEMS will be jointly benchmarked with representative QEC decoders, including minimum-weight perfect matching and neural-network-based decoders, under realistic hardware noise.

\section*{Data availability}
The datasets supporting the findings of this study are available in the Zenodo repository at \url{https://zenodo.org/records/18230210}.
\\
\section*{Code availability}
The custom code used for PAEMS circuit-level simulation, parameter optimization, and correlation analysis is publicly available at
\href{https://github.com/hesonghuan17-netizen/PAEMS-Precise_and_Adaptive_Error_Model_for_superconducting_Quantum_Processors}{\nolinkurl{https://github.com/hesonghuan17-netizen/PAEMS-Precise_and_Adaptive_Error_Model_for_superconducting_Quantum_Processors}}.
% \\

\vspace*{2cm}

% \clearpage
% \newpage
% \noref{*}

\section*{Acknowledgements}
This work was supported by the National Natural Science Foundation of China (Grant No. 62271129) and the Sichuan Provincial Central Government Guiding Local Science and Technology Development Program (Grant No. 2025ZYD0153). This work was partly supported by Tianyan Quantum Computing Coud Platform developed by China Telecom Quantum InfomationTechnology Group Co., Ltd.
\section*{Author contributions}
H.S. and C.Y. conceived the PAEMS framework and designed the overall methodology.
W.C. supervised the project and contributed to the interpretation of results.
L.B. and G.K. provided guidance, assistance, and technical support throughout the project.
All authors discussed the results and contributed to the preparation of the manuscript.
\section*{Competing interests}
The authors declare no competing interests.

\end{document}